\documentstyle[12pt]{article}
\begin{document}

\def\etal{\it et al.\/}
\def\refskip{\vskip 1.0mm}

\title{Dick Dalitz: \\
Examples of His Contributions to Particle Physics}
\author{Gary R. Goldstein\thanks{Supported in part by U.S. Department of Energy grant DE-FG02-92ER40702.}\\
Department of Physics and Astronomy\\
Tufts University\\
Medford, MA  02155 USA\\}
\refskip
\maketitle

I had been asked to speak about the contributions of Dick Dalitz to Particle Physics. I was very pleased to be able to do this as an expression of my good fortune to have worked with him. This was an opportunity to present a synopsis of his contributions to an audience that may not have known about some of this history. In choosing a sampling of Dalitz's work, I relied on the talks that were given at the Memorial Meeting in Oxford, June 1, 2006, the extensive bibliography compiled by Aitchison, Close, Gal, and Millener~\cite{biblio} and my own recollections during many years of collaboration.

Richard Henry Dalitz was born in 1925 in rural Australia, Dimboola, Victoria. His mother was from Scottish forebears. His father was descended from a small, distinctive ethnic group, the Wends or Sorbs. A large contingent of that group had emigrated from Germany to Australia in the mid-19th century. Later in life Dick discovered that he was not of German descent when he uncovered this geneaology. At age two his family moved near to Melbourne, at his mother's urging. He had speculated that perhaps she was conscious of his precocious abilities, even that early, and wanted to be near better schools. This was apparently not an easy change for his father, who managed to earn a modest income as a laborer. In any case, Dick distinguished himself throughout his schooling, culminating in his undergraduate work at University of Melbourne. He was then awarded a traveling scholarship that allowed him to pursue graduate studies at Cambridge University. 

In Cambridge Dick worked with the theorist, Nicholas Kemmer, on nuclear questions, receiving his D. Phil. in 1950. His dissertation was on 0$\rightarrow$0 transitions in $^{16}$O. The gamma transition is forbidden by angular momentum conservation. But the conversion of a virtual longitudinal $\gamma$ into an $e^+ e^-$ pair is allowed and observed. So he encountered pair production early in his career. He soon applied this idea to particle decays, surmising that in electromagnetic decays like $\pi^0\rightarrow \gamma \gamma$ one or both of the $\gamma$'s could appear as an $e^+e^-$ pair, since known as Dalitz pairs. Such pairs were seen in nuclear emulsions soon thereafter. The $\pi^0\rightarrow e^+ e^- e^+ e^-$ decay was used to determine the parity of the $\pi^0$~\cite{pi0} through the azimuthal correlations of the pairs. A similar analysis of $\Sigma^0$ decays into $\Lambda^0+e^+e^-$ showed that the two hyperons had the same parity~\cite{lambda}.

It is interesting to speculate about how outstanding scientists came to their insights; where there creative ideas originated; what habits of mind were instrumental in their important work. With Dick Dalitz there is evidence from this earliest work on nuclear transitions and Dalitz pairs that thorough, detailed studies of the phenomena in question open up the mind for creative leaps. In much of his work he began with experimenters showing him puzzling phenomena or asking difficult questions about poorly understood data. The physics was in the observable phenomena. In this sense he was always a phenomenologist, although he could get quite involved in the subtleties of field theory in many instances. 

While still working on his thesis he spent part of a year visiting the cosmic ray group of C. Powell in Bristol where strange particles were being discovered and studied. He became very interested in hyperfragments and hypernuclei, nuclei with a hyperon replacing a nucleon. He continued to revisit this subject throughout his career, seeing it as a rich source for understanding the many manifestations of the strong, nuclear interaction. I will not discuss his extensive work on hypernuclei here (a summary and references are presented in the aforementioned ref.~\cite{biblio} and by his collaborator, A. Gal in ref.~\cite{gal}). In this period he was drawn into the puzzling properties of strange particles, especially strange mesons that decayed into pions. Two seemingly different strange mesons, the $\theta^+$ and the $\tau^+$, had about the same mass but decayed into two and three $\pi$'s, respectively. For the two $\pi$ decay the angular momentum of the $\theta$ had to be $0^+,1,^-,2^+,$ etc. To determine the angular momentum of the $\tau$ he developed a geometrical analysis. Three particles in the rest frame of the decay have to have their energy add to the mass of the $\tau$ and their 3-momenta add to zero. With each $\pi$'s energy related to its 3-momentum, there are only two independent variables left to characterize any particular decay. The distribution of events in a 2-dimensional plot will reveal the allowed mutual orbital angular momenta. He called this a ``phase space plot", but its remarkable usefulness warranted the eponymous designation as the ``Dalitz plot". When applied to the $\tau$ data it showed that the particle had to have even spin and odd parity,  $0^-, 2^-,$ etc. This pair of nearly equal mass strange particles have opposite parity, a circumstance known then as the $\tau-\theta$ puzzle. He speculated that perhaps there was something amiss with parity conservation, but found little sympathy for this idea among his colleagues. In 1956 T.D. Lee and C.N. Yang wrote a paper in which they concluded that there was no evidence that parity was conserved in weak processes, although it was clearly conserved in strong and electromagnetic processes. This would solve the puzzle - the $\tau$ and $\theta$ are the same particle and parity is not conserved in their decays. By 1957, C-S. Wu and collaborators showed that parity was not conserved in nuclear beta decay, ushering in a new era in particle physics. 
\begin{figure}
\vspace{1.7in}
\includegraphics{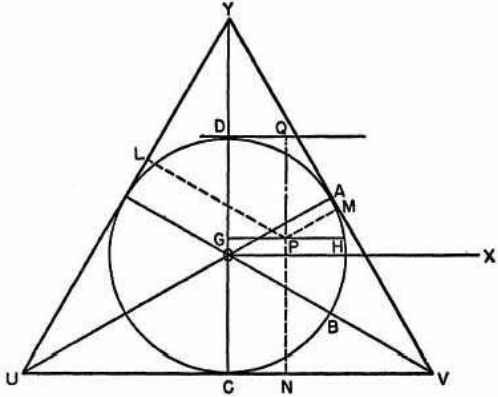}
\vspace{0.3in}
\caption{Triangular diagram for the representation of $\tau$-meson decay events, from Ref.~\cite{plot}.}
\label{plot1}
\end{figure}

\begin{figure}
\vspace{2.5in}
\includegraphics{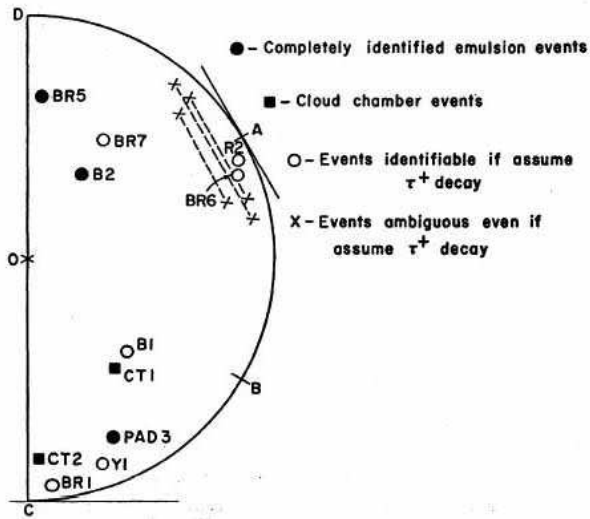}
\vspace{0.3in}
\caption{Data from $\tau$-meson events in which the $\pi$-meson charges are established, from Ref.~\cite{plot}.}
\label{plot2}
\end{figure}

\begin{figure}
\vspace{1.8in}
\includegraphics{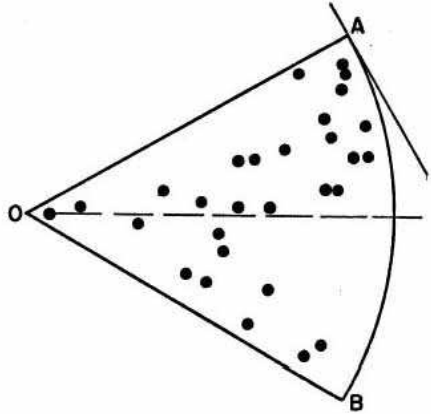}
\vspace{0.3in}
\caption{29 $\tau$-meson decay events from Ref.~\cite{plot}.}
\label{plot3}
\end{figure}

The plots that were displayed in Dalitz's first Physical Review article on the subject~\cite{plot} are shown here in Figs~\ref{plot1},\ref{plot2},\ref{plot3}. Here was a major step forward in analysis and it is applied with great care. 
It is particularly noteworthy that his hand drawn figures have such precision. He respected the importance of data. Throughout his career he valued working closely with experimentalists. There are many instances in which discussions with experimenters and study of current data inspired his theoretical breakthroughs. The Physical Review article also demonstrates the care and precision of his writing, in general. 
His text (and his physics) is never carried away by hyperbole. Clarity of expression was central to his approach to writing and speaking. In the article he states his conclusion cautiously, ``At the present stage, the number of $\tau$-decay events giving a slow unlike $\pi$ meson rather suggests that the least $L$ is $L=0$, which would imply that the $\tau$ meson belongs to a class of even $j$ and odd $w$, a class for which the $2\pi$ decay is forbidden." He awaits further data to confirm this assignment, but sees that it is quite likely, given the distribution of this small number of events. Soon thereafter more data did confirm his observation, $\tau$ was $0^-$.

After leaving Cambridge and then Bristol, where he came to know the cosmic ray data that inspired his interest in strange particles, he became a Lecturer in Birmingham at the invitation of Rudolph Peierls. Peierls became a lifelong colleague and friend. In Birmingham and in visits to several US facilities, he pursued the analyses that expanded on discoveries of strange particle resonances. The Dalitz plot was instrumental in finding the  $\Sigma^*$(1385) in 1960~\cite{sigma} and later the $\Xi^*$(1530). Both had spin 3/2 and are strange partners of the first hadronic resonance, the $\Delta$(1236) discovered by Fermi and Anderson in 1952~\cite{delta}. With various collaborators, Dalitz studied strong production processes of strange particles - mesons and hyperons.  He pursued this research through multichannel partial wave analyses, supplemented by a growing understanding of the analyticity of the scattering amplitudes. Much of this is thoroughly summarized in several review articles and a nicely written 1962 book for workers in the field and graduate students, Strange Particles and Strong Interactions~\cite{book}. 

Of special note in the middle 1950's is Dalitz's prediction, with S-F. Tuan, of a $\Lambda$(1405) hyperon below the $K^- + p$ threshold~\cite{tuan}, which appeared in the partial wave analysis extended below threshold via a multichannel K-matrix method. This state was discovered subsequently, but its complicated interactions have remained an interesting area of study itself. In this period Dalitz probed deeply into the subtleties of partial wave analyticity and, along with Castillejo and Dyson in 1956, found an unnoticed phenomenon, since called the CDD pole~\cite{cdd}. The dispersion relations for amplitudes that arise from strong interaction dynamics in meson-baryon scattering (for example the Low equation) contain resonance poles and threshold cuts. But the implementation of analyticity and unitarity also allows a set of extra poles with arbitrary strengths and positions - the CDD poles. These can be associated with fundamental, elementary particles or with overall ambiguity in the dynamical equations. These poles continue to be of use in understanding many two-body scattering processes.

Dick settled into the University of Chicago from 1956 to 1963, during which there were many experimental discoveries of new meson and baryon resonances. The Dalitz plot had application to many of those discoveries. It was clear, though, that the plethora of particles could not all be fundamental. Theorists looked for patterns that would allow grouping these particles. Dalitz's study of strange states primed him to appreciate efforts to generalize from isospin to more encompassing symmetries. Gell-Mann's and Ne'eman's SU(3) was proposed in 1961 and Dalitz soon studied its impact on the spectrum of strange states. A new notion about universality emerged and to address strangeness changing weak decays, Cabibbo, in 1963, proposed a mixing angle between the strange and non-strange sector. Dick was interested in comparing the mixing angles for pseudoscalar mesons to vector mesons. He also saw tests of universality as experimentally accessible and urged experimenters to test its ramifications~\cite{zichichi}.  When Dick first heard about Gell-Mann's proposal at a 1961 Summer School in India, he was struck by the notion that there could be more fundamental particles, the quarks, even though they were initially only considered to be a mathematical device (triplet representation of flavor SU(3)) for forming group representations. In the next few years, during which he moved to Oxford as the Royal Society Professor (at Peierls' urging), he thought about how to form the observed hadrons out of actual bound physical quarks. By the time of the Les Houches Summer School in 1965~\cite{les} he presented a thoroughly developed quark model of the baryons. 

\begin{figure}
\vspace{2.7in}
\includegraphics{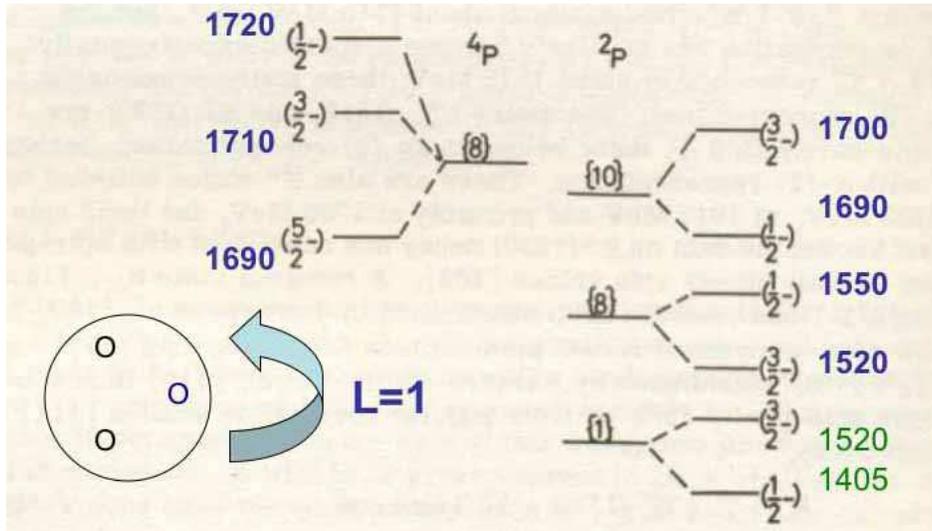}
\vspace{0.3in}
\caption{L=1 multiplet from Ref.~\cite{les} with masses from Ref.~\cite{frank}.}
\label{L1multiplet}
\end{figure}

\begin{figure}
\vspace{2.5in}
\includegraphics{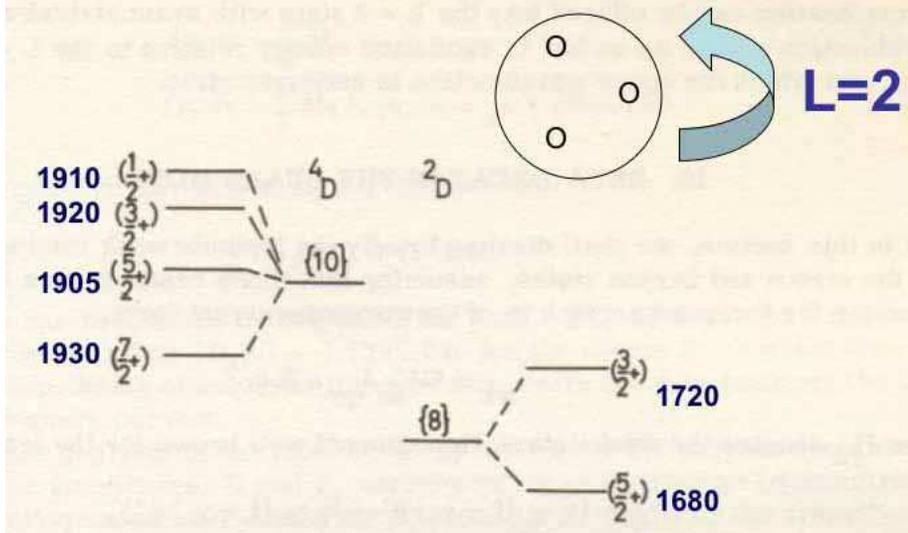}
\vspace{0.3in}
\caption{L=2 multiplet from Ref.~\cite{les} with masses from Ref.~\cite{frank}.}
\label{L2multiplet}
\end{figure}

Dalitz's ``symmetric quark model" is a non-relativistic construction in which 3 quarks are bound by unspecified two-body forces in various orbital angular momentum configurations with spatially symmetric, antisymmetric or mixed symmetry wavefunctions. The overall space-spin-flavor symmetry is symmetric under pair interchanges, thereby violating Fermi statistics. He knew that this was a problem to be solved somehow, but, nevertheless saw the model as incorporating most of the known baryons into well-defined multiplets. The ground state spin $\frac{1}{2}$ octet and spin $\frac{3}{2}$ decuplet were s-states with masses separated by hyperfine splitting. The first excited states were $L=1$ and $L=2$. Fig.~\ref{L1multiplet} and Fig.~\ref{L2multiplet}  show these states along with Frank Close's~\cite{frank} additions of the currently known masses for these states. The connection to current data is remarkable, albeit that 40 years have elapsed. A glance at the current Particle Data Group Listings of baryons shows how remarkable is the reduction of the spectrum to three quark states. 
This model holds up, given the addition of the color SU(3) symmetry introduced several years later that allows the ``symmetric" quark model to become the antisymmetric quark model in compliance with Fermi statistics. While Dick worked on this model, in 1964 the unification of spin and SU(3) had been suggested via SU(6) by Sakita and by G{\" u}rsey and Radicati~\cite{su6}. The Dalitz model ground state baryons then fit nicely into a {\bf 56} representation of SU(6), while the next excited states are in {\bf 70} and {\bf 20} representations.

In the late 1960's through the 1980's Dick continued to work on hadron spectroscopy from the quark model QCD perspective, hadron scattering as described via analyticity and multichannel unitarity, hypernuclei, CP violation and other symmetries in weak decays. Dick even worked on the formulation of potential models in Lattice QCD. He worked with many colleagues, post-docs and students, and built the Theoretical Physics Department at Oxford after R. Peierls stepped down. He also became interested in the biographies of his notable forerunners and peers, writing several scientific biographies, including those of Peierls, Dirac, Skyrme, Tamm and Sakharov.

It was in 1987 that I began collaborating with Dick. R. Marshall, an experimenter at Rutherford Lab and member of a DESY collaboration, had raised a question to Dick. How does a quark spin polarization get transferred to the jet of hadronic fragments that the quark produces? Dick invited me to spend part of my sabbatical in Oxford to address this question and whatever else might arise. We decided that this could be done in principle by considering three of the most energetic mesons in the jet. To identify the particular quark flavor would be difficult unless it were a heavy flavor. At that time there were many charmed quark processes, but few bottom quark mesons and the top quark had not been discovered. So we focused on c-quarks and $D\pi\pi$ fragments. The correlation represented by $\langle \vec{S}\cdot \vec{p}_D\rangle \langle \vec{p}_1\cdot \vec{p}_2\rangle$ can be non-zero for non-zero c-quark polarization $\vec{S}$, either helicity or transversity. Careful examination shows that this correlation arises from an interference between a tree-level amplitude and the imaginary part of an amplitude for final state interactions, a so-called $T$-odd effect (although there is no violation of Time Reversal invariance). We performed a model calculation involving the interference between independent amplitudes for c-quark $\rightarrow$ s-wave $D\pi\pi$ and c-quark $\rightarrow D^* \pi$~\cite{dgm}. Disappointingly, the predicted effect was small, which was confirmed at SLAC. Subsequently fragmentation of transversely polarized light quarks into mesons with transverse momentum was reexamined by J. Collins~\cite{collins} and the prediction and measurement of the ``Collins function" has become a very active pursuit.
 
\begin{figure}
\vspace{2.5in}
\includegraphics{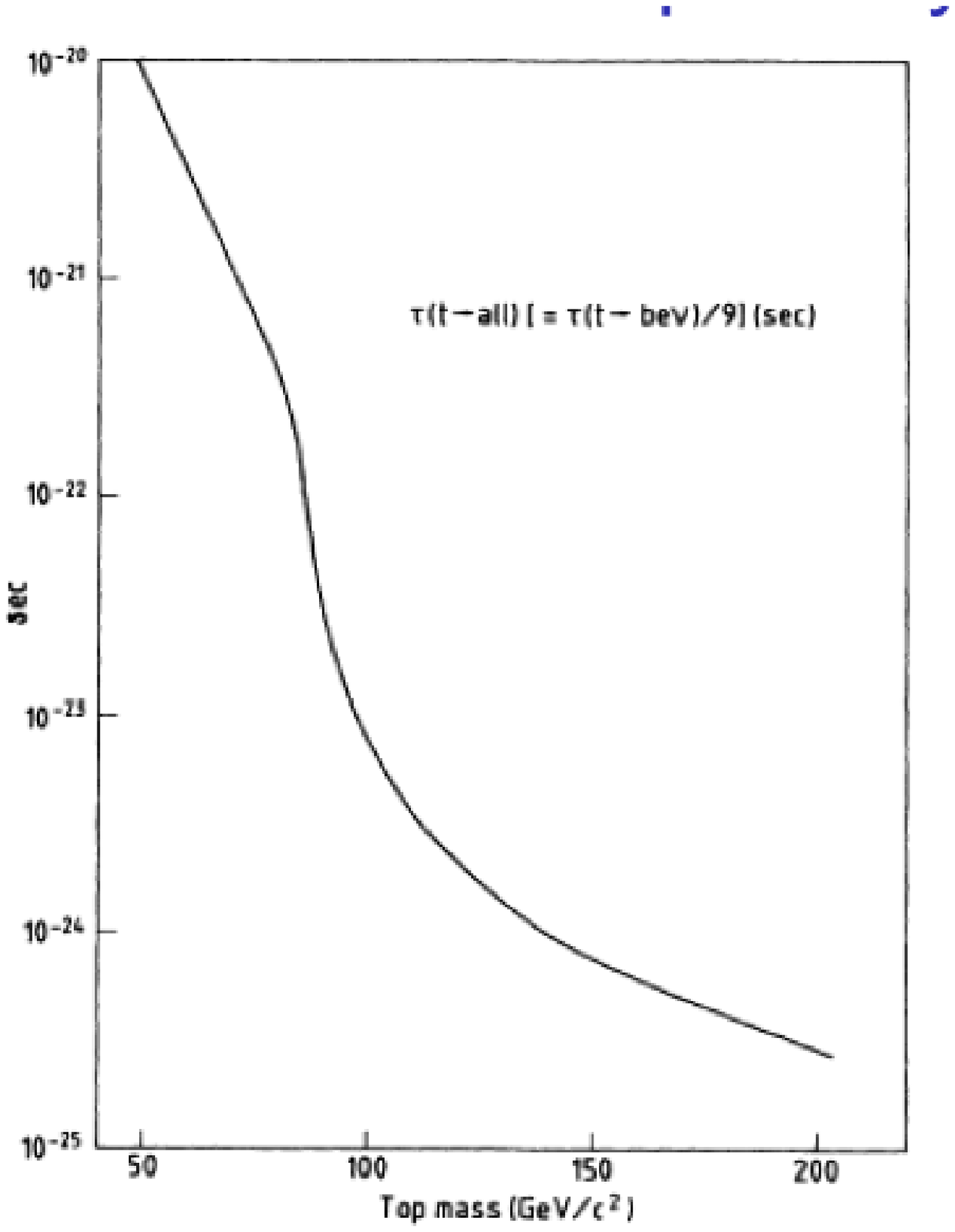}
\vspace{0.3in}
\caption{Top quark lifetime vs. mass from Ref.~\cite{top1}.}
\label{lifetime1}
\end{figure}

\begin{figure}
\vspace{2.8in}
\includegraphics{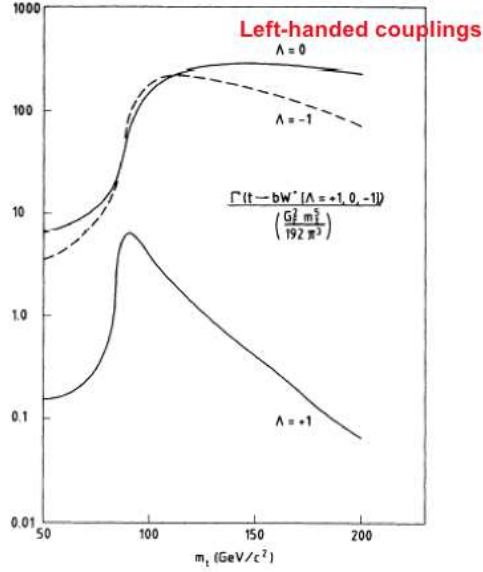}
\vspace{0.2in}
\caption{Partial rates for different helicities  Ref.~\cite{top1}.}
\label{lifetime2}
\end{figure}

In 1989 five quark flavors were well established and the sixth, the top quark was being sought at LEP and Fermilab. Non-discovery at LEP put a lower limit on the top mass of 100 GeV. Dick and I realized that for a mass considerably larger than that, the top would decay very rapidly into a b-quark and an on-shell W boson. The Standard Model calculation showed that the lifetime would go as $1/m_t^5$ and would be comparable to strong decays. Hence, the t-quark would decay as fast or faster than the time it would take to form color singlet, top flavored hadrons, as shown in Fig.~\ref{lifetime1} and~\ref{lifetime2}. 
The possibility of producing tops at LEP inspired our first paper~\cite{top1} on $e^+e^-$ production characteristics. In the clean reaction $e^+e^-\rightarrow t + {\bar t}$, the incoming energy would fix the top mass. Since the mass was too high for LEP, however, it would fall to the Tevatron to search for it. The collider at that time provided near 1 TeV for $\bar{p} p$ energies, but the top mass is no longer tied to the incoming energy.
Furthermore, the most striking signature for the production would involve two missing neutrinos, which would seem to make kinematic reconstruction difficult and full of ambiguity. But this was the kind of challenge that inspired Dick to push forward. We thought about how particular events would have to appear in momentum space, given the chains of subsequent decays. A very attractive geometrical picture emerged~\cite{top2}. We used the Standard Model to determine the probability $P_i(m_t)$ that in an event $i$ a measured set of jet and lepton momenta corresponds to top quark pair production and decay for top quark mass $m_t$. Underlying this probability is the interpretation that an observed event is consistent with the hypothesis
$$\bar{p} + p \rightarrow \bar{t} + t + {\rm other\: hadrons}$$

$${\rm followed \: by \: \: \:} t \rightarrow W^{+} b $$

$${\rm where\: \: \:}
W^+ \rightarrow l^+\nu_l \:{\rm or}\: u\bar{d} \:{\rm or}
\: c\bar{s} $$

$${\rm and \: \: \: }
\bar{t} \rightarrow W^{-} \bar{b} $$

$${\rm where \: \: \:}
W^- \rightarrow l^-\bar{\nu}_l \:{\rm or}\: d\bar{u}\:
{\rm or}\: s\bar{c}.$$
Given that the $W$ decay into quarks involves pairs of colored quarks, the quark jets are about 9 times more likely than the lepton mode. Nevertheless, the pair of leptonic decays produces very energetic, large angle leptons that signal a large mass origin. In this dilepton channel the probability depends critically on the assumed mass of the top quark, since there are two unobserved neutrinos. In the unilepton channel~\cite{top3} the probability is more constrained, since the purely hadronic decay of $t \rightarrow b + q +\bar{q}'$ should give an invariant mass for the decaying top. The invariant mass is smeared, however, by the hadronization process as well as experimental uncertainties in momentum assignments for jets.

The geometrical construction that yields a probability distribution for the top mass is based on 4-momentum conservation, as follows. Consider one hypothetically produced top, with unknown mass $m_t$, energy $E_t$ and 3-momentum $\vec{t}$. The possible $\vec{t}$ values are plotted in 3-dimensional momentum space. With decay products' energy and momenta labeled by $b, l, \nu$, the 3-momenta and energies are constrained via 

$\left(\vec{t} - \vec{b}\right)^2 = \vec{W}^2,$ a sphere in $\vec{t}$-space centered on $\vec{b}$,
 
$\left(\vec{t} - \vec{b}-\vec{l}\right)^2 = \vec{\nu}^2$, a sphere centered on $\vec{b}+\vec{l}$,

$\vec{W}^2=\left(E_t-E_b\right)^2 - M_W^2 $, fixing the radius of the first sphere,

$E_{\nu}^2=\left(E_t-E_b-E_{l}\right)^2$, fixing the radius of the second sphere.

\begin{figure}
\vspace{2in}
\includegraphics{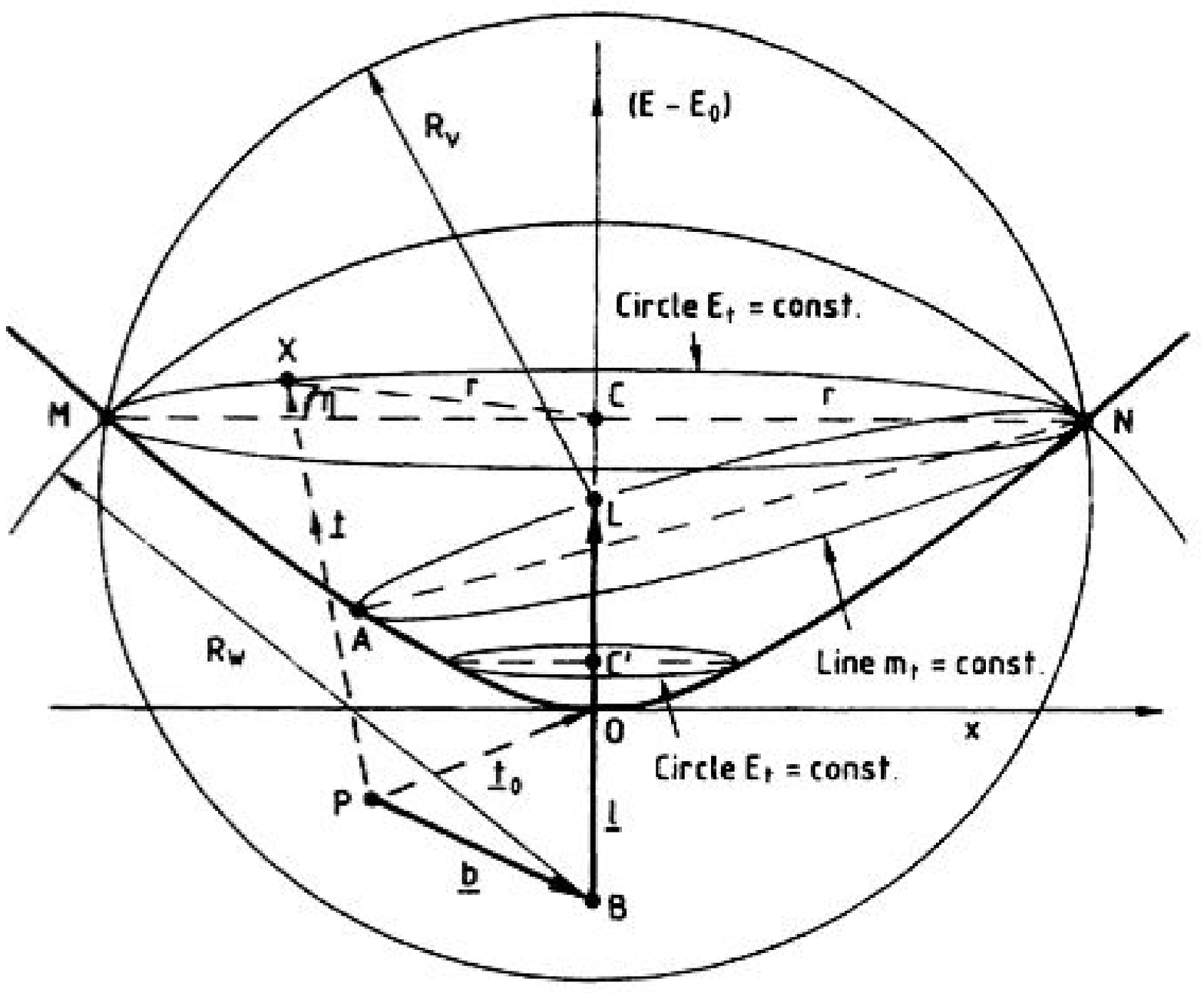}
\vspace{0.3in}
\caption{Construction of allowed top momenta $\vec{t}$ for given $b$ and $l$ momenta, see Ref.~\cite{top2}.}
\label{sphere}
\end{figure}

\begin{figure}
\vspace{2.5in}
\includegraphics{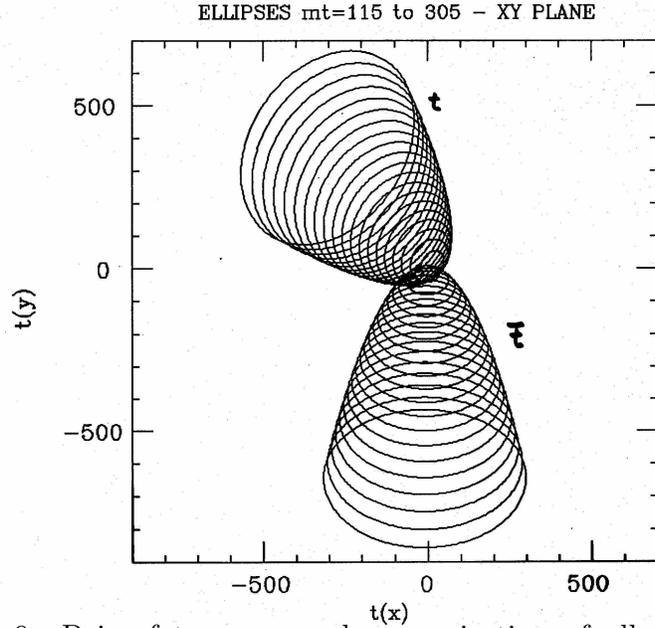}
\vspace{0.3in}
\caption{Pair of transverse plane projection of allowed top and antitop momenta.}
\label{ellipses}
\end{figure}

For fixed $t$ 4-momentum, the $\vec{t}$ lies on the circle of intersection of the two spheres. The circle lies in a plane perpendicular to the $\vec{l}$. The set of all $E_t$'s forms a parabaloid with axis along $\vec{l}$ as shown in Fig.~\ref{sphere}. For fixed $m_t$, $E_t^2- \vec{t}^2$ is constrained, cutting the paraboloid in an ellipse. The collection of ellipses for all allowed $m_t$ values covers the paraboloid. A corresponding set of ellipses for the anti-top can be constructed also. The quark+antiquark pair in each event that annihilate to form the top+antitop pair have limited transverse momenta, so that $\vec{t}_T + \vec{\bar{t}}_T \simeq 0$. 
Hence the intersections of each pair of fixed $m_t$ ellipses determine possible values of the top and antitop momenta, as in the simulated example of Fig.~\ref{ellipses}. At each intersecting pair of momenta for a possible $m_t$ there is a Standard Model probability assigned. So each event gets a distribution of probabilities as a function of top mass.

As candidate events became available our method was applied to determine the relative probability that an event involved top production. For likely cases the probability distribution as a function of possible mass yielded peak values. Collecting the first sizable set of likely top events and their peak mass probabilities gave us an estimate of the actual top mass. This value, around 160 GeV with a big spread, was on the low side of the currently reported mass nearer to 174 GeV~\cite{top4}. More recently a variant of this method was used by the D0 group to further constrain the top mass and to set limits on the Higgs mass~\cite{d0}.

Dick Dalitz continued to work on hypernuclei and scientific biographies into the last years. His pace slowed, but his enthusiasm for interesting problems never waned, nor did the seriousness with which he approached his work. He was an active participant in Particle Physics during its most exciting years of discovery. He contributed significantly to those discoveries. Dick's deep, sincere, overarching appreciation of physics as a pursuit of knowledge and understanding, without cant or boasting, was genuine and rare. He never expressed doubt that this pursuit was of unquestionable value and importance. Those who knew Dick Dalitz and those who were affected by his work will miss him. His dedication to physics was a beacon for us all. 

\vspace{0.5in}

\noindent{\bf Acknowledgement}

\noindent It is a pleasure to thank Prof. Zichichi for inviting me to the Erice meeting and to thank the staff that makes this Summer School so successful. This is a truly remarkable institution whose longevity and reputation are well deserved.

\end{document}